\documentstyle[11pt,psfig,twoside]{article}

\begin{document}

\newcommand{\dd}{\,{\rm d}}
\newcommand{\ie}{{\it i.e.},\,}
\newcommand{\etal}{{\it et al.\ }}
\newcommand{\eg}{{\it e.g.},\,}
\newcommand{\cf}{{\it cf.\ }}
\newcommand{\vs}{{\it vs.\ }}
\newcommand{\zdot}{\makebox[0pt][l]{.}}
\newcommand{\up}[1]{\ifmmode^{\rm #1}\else$^{\rm #1}$\fi}
\newcommand{\dn}[1]{\ifmmode_{\rm #1}\else$_{\rm #1}$\fi}
\newcommand{\upd}{\up{d}}
\newcommand{\uph}{\up{h}}
\newcommand{\upm}{\up{m}}
\newcommand{\ups}{\up{s}}
\newcommand{\arcd}{\ifmmode^{\circ}\else$^{\circ}$\fi}
\newcommand{\arcm}{\ifmmode{'}\else$'$\fi}
\newcommand{\arcs}{\ifmmode{''}\else$''$\fi}
\newcommand{\MS}{{\rm M}\ifmmode_{\odot}\else$_{\odot}$\fi}
\newcommand{\RS}{{\rm R}\ifmmode_{\odot}\else$_{\odot}$\fi}
\newcommand{\LS}{{\rm L}\ifmmode_{\odot}\else$_{\odot}$\fi}

\newcommand{\Abstract}[2]{{\footnotesize\begin{center}ABSTRACT\end{center}
\vspace{1mm}\par#1\par
\noindent
{~}{\it #2}}}

\newcommand{\TabCap}[2]{\begin{center}\parbox[t]{#1}{\begin{center}
  \small {\spaceskip 2pt plus 1pt minus 1pt T a b l e}
  \refstepcounter{table}\thetable \\[2mm]
  \footnotesize #2 \end{center}}\end{center}}

\newcommand{\TableSep}[2]{\begin{table}[p]\vspace{#1}
\TabCap{#2}\end{table}}

\newcommand{\FigCap}[1]{\footnotesize\par\noindent Fig.\  %
  \refstepcounter{figure}\thefigure. #1\par}

\newcommand{\TableFont}{\footnotesize}
\newcommand{\TableFontIt}{\ttit}
\newcommand{\SetTableFont}[1]{\renewcommand{\TableFont}{#1}}

\newcommand{\MakeTable}[4]{\begin{table}[htb]\TabCap{#2}{#3}
  \begin{center} \TableFont \begin{tabular}{#1} #4 
  \end{tabular}\end{center}\end{table}}

\newcommand{\MakeTableSep}[4]{\begin{table}[p]\TabCap{#2}{#3}
  \begin{center} \TableFont \begin{tabular}{#1} #4 
  \end{tabular}\end{center}\end{table}}

\newenvironment{references}%
{
\footnotesize \frenchspacing
\renewcommand{\thesection}{}
\renewcommand{\in}{{\rm in }}
\renewcommand{\AA}{Astron.\ Astrophys.}
\newcommand{\AAS}{Astron.~Astrophys.~Suppl.~Ser.}
\newcommand{\ApJ}{Astrophys.\ J.}
\newcommand{\ApJS}{Astrophys.\ J.~Suppl.~Ser.}
\newcommand{\ApJL}{Astrophys.\ J.~Letters}
\newcommand{\AJ}{Astron.\ J.}
\newcommand{\IBVS}{IBVS}
\newcommand{\PASP}{P.A.S.P.}
\newcommand{\Acta}{Acta Astron.}
\newcommand{\MNRAS}{MNRAS}
\renewcommand{\and}{{\rm and }}
\section{{\rm REFERENCES}}
\sloppy \hyphenpenalty10000
\begin{list}{}{\leftmargin1cm\listparindent-1cm
\itemindent\listparindent\parsep0pt\itemsep0pt}}%
{\end{list}\vspace{2mm}}

\def\TYLDA{~}
\newlength{\DW}
\settowidth{\DW}{0}
\newcommand{\dw}{\hspace{\DW}}

\newcommand{\refitem}[5]{\item[]{#1} #2%
\def\REFARG{#3}\ifx\REFARG\TYLDA\else, {\it#3}\fi
\def\REFARG{#4}\ifx\REFARG\TYLDA\else, {\bf#4}\fi
\def\REFARG{#5}\ifx\REFARG\TYLDA\else, {#5}\fi.}

\newcommand{\Section}[1]{\section{#1}}
\newcommand{\Subsection}[1]{\subsection{#1}}
\newcommand{\Acknow}[1]{\par\vspace{5mm}{\bf Acknowledgments.} #1}
\pagestyle{myheadings}

\def\thefootnote{\fnsymbol{footnote}}

\begin{center}
{\Large\bf Searching for Variable Stars in the Central Part of the 
Globular Cluster M22}
\vskip1cm
{\bf
Pawe\l\ ~~P~i~e~t~r~u~k~o~w~i~c~z and Janusz ~~K~a~l~u~z~n~y}
\vskip3mm
{Copernicus Astronomical Center, Bartycka 18, 00-716 Warsaw, Poland\\
e-mail: (pietruk,jka)@camk.edu.pl}
\end{center}

\Abstract{

Time-series data taken with the Hubble Space Telescope of three 
fields covering the central part of the globular cluster M22
have been analyzed in search of variable objects.
We report identification of 11 periodic variables of which 8 are new
discoveries. The sample includes 5 certain contact binaries
as well as 1 SX~Phe star.
Two objects with periods longer than 1 day are preliminarily
classified as either spotted variables of BY Dra type or ellipsoidal 
variables.
The most unusual of the identified variables, M22\_V11, has $I_C\approx 19.9$
and is located far to the red of the main sequence in 
the cluster color-magnitude diagram. It shows variability with a 
period as short as $P\approx 0.066$~days or alternatively
with $P\approx 0.132$~days.
We propose that it may be an ellipsoidal variable harboring a degenerate 
component.
}
{globular clusters: individual: M22 -- binaries: eclipsing --
-- stars: variables}

\Section{Introduction}

M22 (NGC 6656) is a bright and rich globular cluster projected
against the outer part of the Galactic bulge. 
Clement {\it et al.} (2001) list 79 variable stars known to date in the 
cluster field. Most of them are RR Lyr stars or red long-period variables.
Several SX Phe stars and candidate eclipsing binaries
were recently identified by Kaluzny \& Thompson (2001).

The cluster is located about one-third of the   
way between the Sun and the Galactic bulge. This makes it
an interesting target for observations aimed at detection
of microlensing events. Recently Sahu {\it et al.} (2001) presented
results of such a survey which was made in 1999 with the
HST/WFPC2. All but one candidate event reported in that paper were,
however, later on dismissed as spurious detections triggered
by cosmic rays (Sahu {\it et al.} 2002).

\Section{Observations and Data Reduction}

The analyzed data were taken by the HST with the WFPC2 camera 
starting on 1999 February 22 and ending on 1999 June 15, as a part of 
the GO 7615 program.  
Three fields covering the central region of M22 were observed.
Their location is shown in Fig. 1. Among 43 observations available for 
each field 31 were made in the F814W filter and 12 were made in
the F606W filter. For both filters every individual observation consist
of a pair of adjacent 260~s exposures. Our search for variable objects 
was based solely on the F814W images for which average time resolution 
equals to about 3.6 days.

Data reduction were performed using the HSTphot stellar photometry
package (Dolphin 2000a, Dolphin 2000b). Before extracting photometry
we performed some image processing steps, like masking bad pixels 
and cosmic-ray rejection.
The latter step was based on comparison of the two adjacent
images each forming an individual observation. The profile photometry 
was obtained  using a library of model point-spread functions 
PSFs supplied along with the HSTphot package. The code was run with 
the "option" parameter set to 512.  It was found, after some tests,  
that such a selection allows to obtain most reliable results in comparison 
with other possible values of "option".
Photometry for the F814W filter was transformed to the standard $I_{C}$
system using the utility available in the HSTphot package.

Before constructing light curves for all detected stars
we performed elimination of some potentially poor measurements which could 
lead to detection of spurious variability during subsequent analysis. 
First we removed all objects
with $\chi>7.0$ and $|sharpness|>0.4$ (see Dolphin 2000a for definition 
of these two parameters).
As the next step we prepared for each field a set of masks
covering images of badly saturated stars. The shape of each mask consists 
of a central disk and a cross filling diffraction spikes.
Objects located in regions covered by these masks were
removed from further considerations. For each field about 20\% of objects
detected by HSTphot were eliminated as a result of two above described steps.

Subsequently, we selected a reference image for each of the
six filter and field combinations and calculated 
the median value of the magnitude offset for each individual frame 
in respect to the reference image. These offset corrections amounted 
in all cases to values smaller than 0.02 mag.
Finally we constructed light curves for all stars
retained on lists corresponding to the reference images. 
The overall quality of photometry derived for the $I_{C}$-band 
is illustrated in Fig. 2. In that figure we plot the rms deviation versus 
average magnitude for the light curves of objects from the sub-field B.

Light curves containing at least 24 data points were analyzed
in a search for variable stars. The search was performed with the 
TATRY code using the multi-harmonic periodogram of 
Schwarzenberg-Czerny (1996). Periodograms were calculated 
for periods ranging from 0.01 to 100 days. The later period 
corresponds to time span of observations. After accounting for 
some overlaps between observed fields the total number of analyzed 
objects amounts to 44800. For each of three surveyed fields
light curves of about 300 most probable candidates for variables
selected with the TATRY were examined by eye.

\Section{Results}

Our search for variable stars led to detection of 11 certain variables.
Their equatorial as well as rectangular coordinates measured in 
the reference images are listed in Table 1. 
Three out of eleven detected variables, M22\_01, M22\_02 and M22\_03
were previously found by Kaluzny and Thompson (2001)
and they are listed in their paper as V27, V13 and V43, respectively.
We note that the surveyed field includes also some other known variables.
However, images of these relatively bright stars were saturated
on the analyzed WFPC2 frames. Moreover, we have not recovered
a candidate cataclysmic variable detected originally
by Sahu {\it et al.} (2001) and discussed more
recently by Anderson {\it et al.} (2003). On several frames
images of that object were affected by saturated pixels and as a result its
light curve obtained by us contained too few data points to be included 
in the sample of objects examined for variability.    

Table 2 gives some basic photometric data on all identified variables. 
Instrumental magnitudes F606W at maximum light are estimated from
phased light curves including at most 12 data points. 
Proposed classification of variables is given in the last column of Table 2. 
Their phased as well as time-domain $I_{C}$ light curves  are 
presented in Fig. 3. Fig. 4 shows the location of variables on the 
$I_{C}$ versus F606W-F814W color-magnitude diagram of the cluster.  

Light curves and  periods of stars \#2-6 are typical for W UMa-type 
contact binaries. Also their positions on the color-magnitude diagram 
are consistent with such a classification. We note parenthetically 
that star \#3 exhibits total eclipse. Variables \#7-8  show periods 
falling into the range occupied by contact binaries. 
However, their light curves show minima too wide
for W UMa-type stars. Moreover, star \#7 is located
to the blue of the cluster main sequence, which would be unusual for
a contact binary belonging to the cluster. At present, secure 
classification of variables \#7-8 is difficult but we note that their
periods may be in fact twice as short as those listed in Table 2. 
Objects \#9-10 can be classified either as BY Dra stars or as ellipsoidal    
variables. In the latter case their periods would be twice as long as the 
values given in Table 2. In such a case phased light curves of the
variables would show two maxima and would resemble light curves of
ellipsoidal variables.

Variable \#11 is potentially the most interesting of all variables 
reported in this contribution. It shows a sine-like light curve with 
an amplitude of about 0.16 mag and a period of $P=0.066$~days.
This type of variability would be consistent with classifying the variable as 
a pulsating SX Phe star. However, SX Phe stars are always located among 
blue stragglers on color-magnitude diagrams of their parent 
clusters. As seen in Fig. 4, variable \#11 is too 
faint and too red to be classified as an SX Phe star belonging to M22.
We have considered the hypothesis that this variable is in fact a highly
reddened background SX~Phe star. The cluster reddening is estimated 
at $E(B-V)=0.34$ (Harris 1996), while according to Schlegel {\it et al.} (1998) 
the total Galactic reddening in the cluster direction amounts to  $E(B-V)=0.33$.
Comparison of these numbers indicates that any star located in the cluster
field can hardly be more reddened than the cluster itself. Hence, the red
color of star \#11 is its intrinsic property and therefore it cannot be
classified as candidate SX~Phe star.  The observed period of
the variable falls within the range of orbital periods observed for cataclysmic 
variables (CVs). However, its red color is rather unusual for an active CV. 
We propose that star \#11 is in fact an ellipsoidal variable
composed of a red dwarf and a compact low-luminosity companion.
An example of such type of objects are "hibernating" cataclysmic 
variables, progenitors of millisecond pulsars or X-ray transients harboring 
millisecond pulsars.  
A separate issue is cluster membership status of the variable.
As can be seen in Fig. 4, \#11 occupies a position far to the red
or alternatively far above main sequence of the cluster. 
That suggests that it is just a foreground object not related to the 
cluster. On the other hand the example of J1740-5340 in NGC 6397 
({\it e.g.}, Ferraro {\it et al.} 2001) tells us that optical companions
to millisecond pulsars can be located to the red of the main sequence
on the color-magnitude diagrams of their parent clusters.
In that context, it would be worth to check the membership status of \#11
by determining its proper motion relatively to the bulk of M22 stars.
As was demonstrated recently by Anderson {\it et al.} (2003),
the available archival HST/WFPC2 data make such a determination
feasible for a fraction of the M22 core. Unfortunately, examination
of the HST/WFPC2 archive shows that for the moment it contains only
two sets of images including M22\_V11. Besides the 1999 images analyzed
in this paper, the variable can be located on a set of 26~s images
collected in 2000. The shallowness of the 2000 data as well as the short
time base of two available epochs makes determination of the relative
proper motion of M22\_V11 problematic.

{\small
\begin{table}
\begin{center}
\caption{Equatorial coordinates and ($X,Y$) positions of variables
on the HST/WFPC2 images}

\vspace{0.4cm}

\begin{tabular}{lccccccc}
\hline
Name & RA(2000.0) & Dec(2000.0) & Field & Chip & Location & Dataset name \\
 & [h:m:s] & [$^{\circ}:':''$] & & & ($X,Y$) & \\
\hline
M22\_01 & 18:36:22.49 & -23:55:12.9 & C & PC1 & (557,778) & u5330105r \\
M22\_02 & 18:36:30.81 & -23:53:46.1 & A & WF3 & (121,527) & u5330101r \\
M22\_03 & 18:36:24.21 & -23:56:19.4 & C & WF4 & (336,630) & u5330105r \\
M22\_04 & 18:36:22.79 & -23:52:48.3 & B & WF2 & (230,637) & u5330103r \\
M22\_05 & 18:36:22.26 & -23:54:32.9 & C & WF2 & (669, 78) & u5330105r \\
M22\_06 & 18:36:25.19 & -23:54:37.3 & C & WF2 & (331,308) & u5330105r \\
M22\_07 & 18:36:26.89 & -23:53:43.3 & A & WF4 & ( 97,290) & u5330101r \\
M22\_08 & 18:36:24.17 & -23:54.10.1 & C & WF2 & (616,425) & u5330105r \\
M22\_09 & 18:36:29.99 & -23:55:42.8 & C & WF3 & (246,698) & u5330105r \\   
M22\_10 & 18:36:27.12 & -23:52:59.7 & A & WF2 & (262,137) & u5330101r \\
M22\_11 & 18:36:30.32 & -23:55:23.0 & C & WF3 & (426,601) & u5330105r \\
\hline
\end{tabular}
\end{center}
\end{table}
}

{\small
\begin{table}
\begin{center}
\caption{Photometric data for the M22 variables}

\vspace{0.4cm}

\begin{tabular}{lccclll}
\hline
Name & $I_{C,max}$ & $\Delta I_C$ & F606W$_{max}$ & $P$ [d]
 & d$P$ [d] & Type \\
\hline
M22\_01 & 16.23 & 0.07 &   -   & 0.042171 & 0.000002 & SX~Phe \\
M22\_02 & 16.46 & 0.41 & 17.36 & 0.281699 & 0.000012 & EW \\
M22\_03 & 17.37 & 0.38 & 18.31 & 0.220502 & 0.000005 & EW \\
M22\_04 & 18.46 & 0.32 & 19.77 & 0.312263 & 0.000016 & EW \\
M22\_05 & 17.25 & 0.19 & 18.15 & 0.242792 & 0.000007 & EW \\
M22\_06 & 17.14 & 0.47 & 17.98 & 0.239431 & 0.000005 & EW \\
M22\_07 & 17.87 & 0.11 & 18.50 & 0.35568  & 0.00005  & Ell/BY \\
M22\_08 & 18.70 & 0.20 & 19.74 & 0.32725  & 0.00003  & Ell/BY \\
M22\_09 & 17.93 & 0.16 & 18.90 & 1.4791   & 0.0007   & Ell/BY \\
M22\_10 & 17.32 & 0.32 & 18.47 & 5.250    & 0.003    & Ell/BY \\
M22\_11 & 19.85 & 0.16 & 22.14 & 0.066226 & 0.000005 & Ell(?) \\
\hline
\end{tabular}
\end{center}
\end{table}
}

\Section{Conclusions}

Our search for variable stars in the globular cluster M22  
leads to the detection of 11 certain periodic variables. 
A small number of observations, consisting of just 31 magnitudes
spread over almost a 4 month period, indicates that our sample 
is highly incomplete. In fact, all identified objects show continuous 
variability which makes their detection far easier than detection of
variables with a short duty cycle, such as detached eclipsing binaries.
Considering the small number of available observations, we decided that 
it was pointless to conduct any simulations aimed at deriving 
quantitative estimates of completness of derived sample of variables.

Results of the astrometric study conducted by Anderson {\it et al.} (2003) 
indicate that inside the surveyed HST/WFPC2 field, the cluster stars 
prevail strongly over field objects. That suggests that most of
detected variables belong to M22. However, dedicated astrometric 
reductions extending the results of Anderson {\it et al.} (2003)
to the remaining 3 WFPC2 chips would be needed to obtain individual
membership probabilities for each of the detected variables.

\Acknow{

We would like to thank Alex Schwarzenberg-Czerny and
Grzegorz Poj-ma\'nski for providing some useful software which was
used in this project. We are grateful to Andrew Dolphin
for helpful hints on the HSTphot package. It is also a pleasure
to thank Stefan Mochnacki for remarks on the draft version of this
paper.

This work is based on observations with the NASA/ESA Hubble Space
Telescope, obtained from the Data Archive at the Space
Telescope Science Institute, which is operated by the Association
of Universities for Research in Astronomy, Inc., under
NASA contract NAS 5-26555. These observations are associated
with program \#7615.

JK was supported by the Polish KBN grant 5P03D004.21.
}

\begin{figure}[htb]
\hglue-0.5cm\psfig{figure=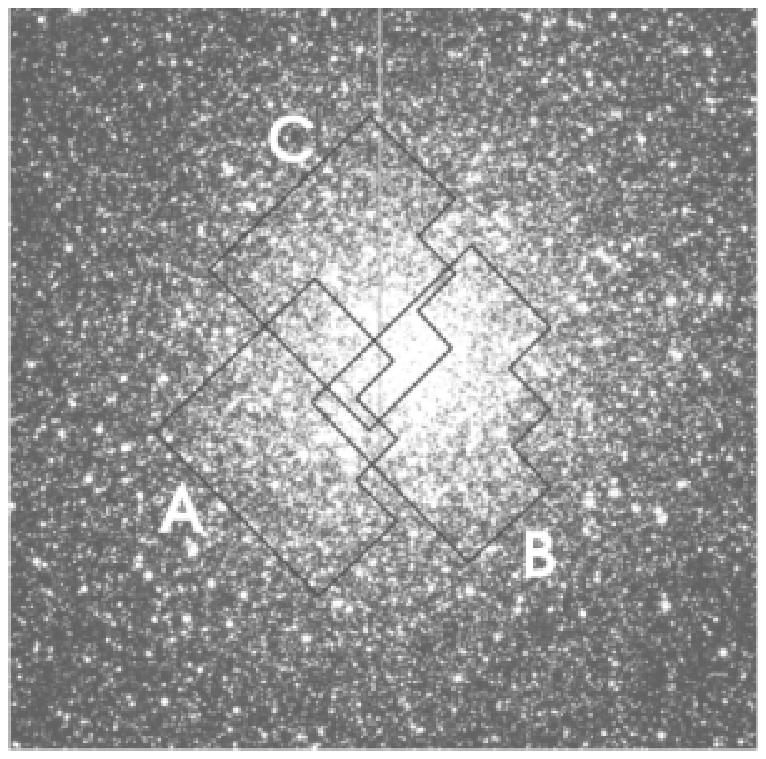,width=13.5cm,angle=0}
\FigCap{
The three HST fields in M22 overlaid on the ground-based image. 
East is to the left and South is up.
}
\end{figure}

\begin{figure}[htb]
\hglue-0.5cm\psfig{figure=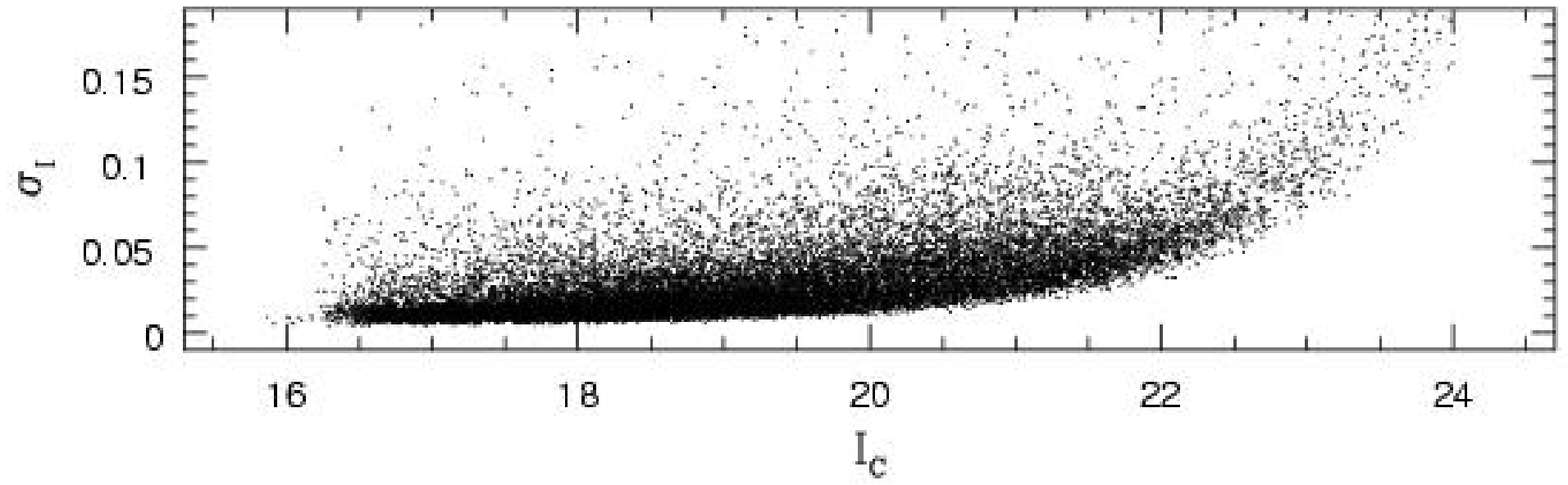,width=13.0cm,angle=0}
\FigCap{
The $rms$ for stars from field B as a function of $I_C$ magnitude.
}
\end{figure}

\begin{figure}[htb]
\hglue-0.5cm\psfig{figure=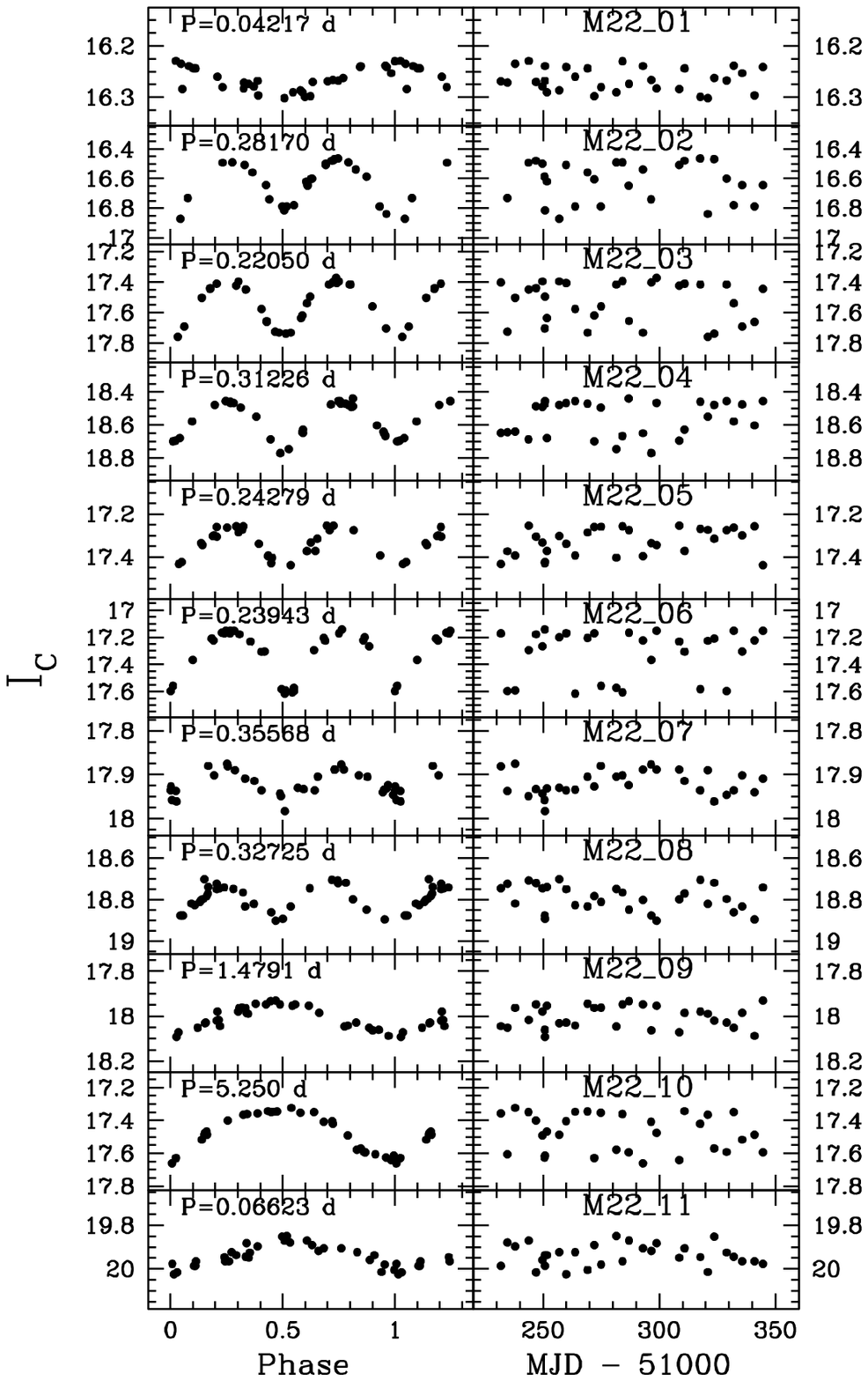,width=13.0cm,angle=0}
\FigCap{
Phased and time-domain (right panel) $I$-band light curves for the 11
detected variables in the field of M22.
}
\end{figure}

\begin{figure}[htb]
\hglue-0.5cm\psfig{figure=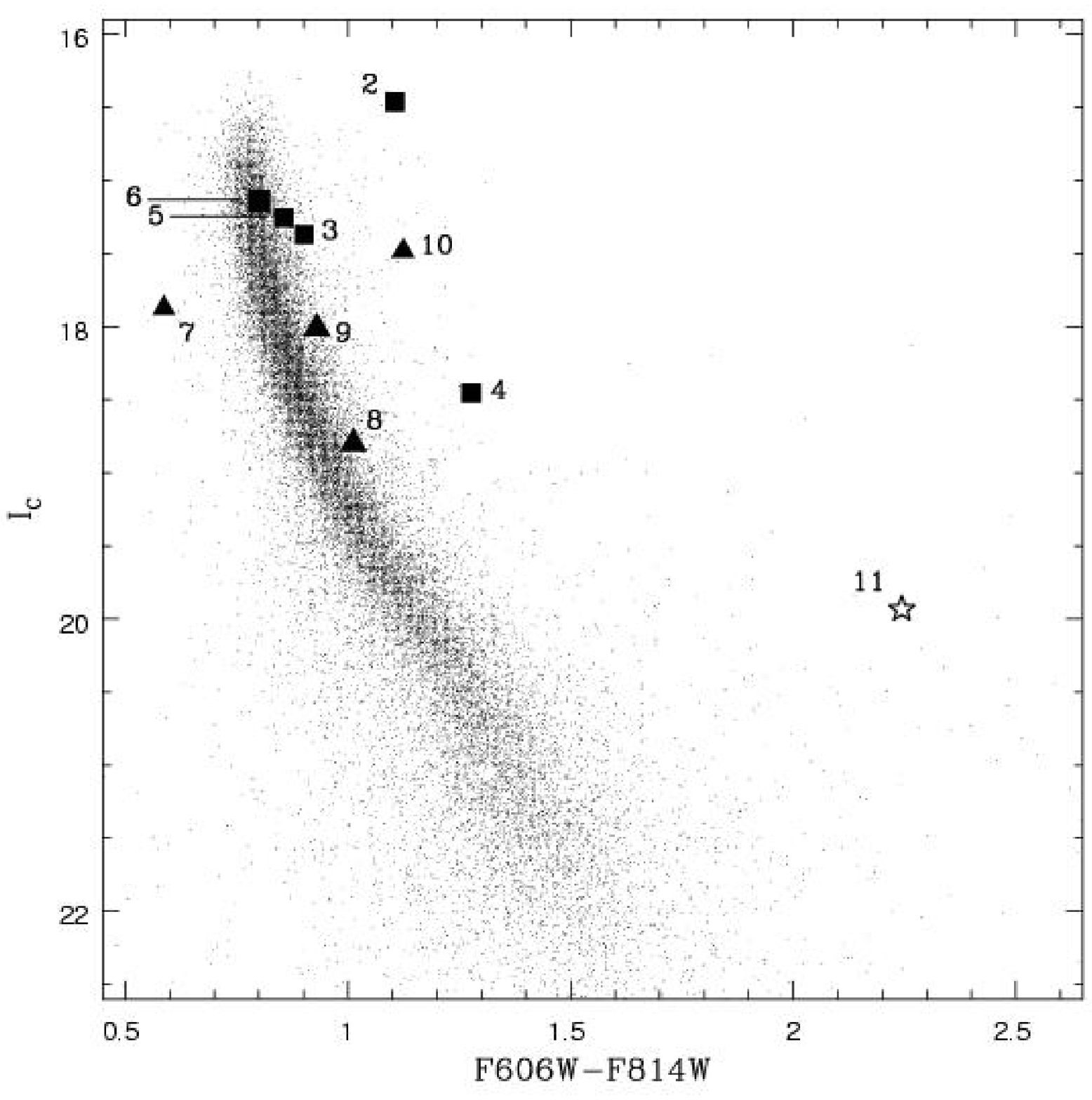,width=13.0cm,angle=0}
\FigCap{
Location of detected variables in the color-magnitude diagram
of M22 (field B) for all four WFPC2 detectors.
Squares denote EW systems.
}
\end{figure}  

\end{document}